\PassOptionsToPackage{square,comma,numbers,sort&compress,super}{natbib}
\documentclass[aps,pre,preprint,superscriptaddress,floatfix,longbibliography]{revtex4}
\usepackage{amsmath}
\usepackage{amssymb}
\usepackage{graphicx}
\usepackage{color}
\usepackage{bm}
\usepackage{textcomp} 
\usepackage{enumitem}
\usepackage{achemso}
\setkeys{acs}{articletitle=true}
\setlist{noitemsep,leftmargin=*}
\usepackage{hyperref}
\hypersetup{
colorlinks=true,
urlcolor= blue,
citecolor=blue,
linkcolor= blue,
bookmarks=true,
bookmarksopen=false,
}

\begin{document}

\title{Non Markovian dynamics induced electric field from a time-independent magnetic field}
\author{Joydip Das}
\affiliation{Department of Chemistry, Visva-Bharati, Santiniketan, India, 731235}
\author{Mousumi Biswas}
\affiliation{Department of Chemistry, Visva-Bharati, Santiniketan, India, 731235}
\author{Debasish Mondal}
\affiliation{Department of Chemistry and Center for Molecular and Optical Sciences \& Technologies, Indian Institute of Technology Tirupati, Tirupati, India, 517 506}	
\author{Bidhan~Chandra~Bag}
	\email{bidhanchandra.bag@visva-bharati.ac.in}
	\affiliation{Department of Chemistry, Visva-Bharati, Santiniketan, India, 731235}	
\begin{abstract}

Using the property of the solution of the Langevin dynamics with a generalized frictional memory kernel
and time-dependent deterministic force field (TDFF) we show that a solution method (which is very simple as well as shortcut) can be used  to derive the Fokker-Planck equation (FPE) for this dynamics. Then the equation is derived for the non-Markovian dynamics with additional force from harmonic potential, magnetic fields, and both, respectively. Thus the method is instructive in deriving the Fokker-Planck equation in a shortcut way in the presence of an additional time-dependent stochastic force. Here we have to consider that the relevant drift terms are independent of the random force. Then another very important point is to be noted here. To interpret FPE, we recognize that
the memory of the non-Markovian dynamics can induce an electric field from the time-independent magnetic field in the presence of a conservative force field. Then one may notice that how it may modulate  diffusion terms and the effect from a time-dependent external force field. 
We understand that the present study, with the recognization of the induced electric field will bring strong attention to different areas of non-equilibrium statistical mechanics such as physical tuning of conductivity of ions in solid electrolytes, noise-induced transition and stochastic thermodynamics for non-Markovian dynamics etc. 
\end{abstract}



\maketitle
\newpage
\section{Introduction}
The Langevin equation of motion is the expected law of causality for the Brownian motion. Its equivalent description may be the Fokker-Plank equation (FPE) for special cases like the Gaussian random force driven Brownian motion. The FPE equation is unavoidable in calculating the barrier crossing rate constant\cite{kram}, the entropy production for the irreversible processes\cite{nico,oth1,baurapre,izumida} and  in the context of the linear response theory\cite{risken} etc. Kramers' theory \cite{kram} on the barrier crossing dynamics requires the relevant probability density function (PDF) at the steady-state. To determine this, we must need the FPE\cite{kram}. In this context, we cite a very recent study, \cite{abdoli} where escape dynamics in a two-temperature Brownian magneto-system is investigated.
Similarly, the Fokker-Planck operator plays an important role in the linear response theory, as demonstrated in Ref.\cite{risken}. Thus finding  the FPE corresponding to a Langevin equation and its use of it are always intriguing issues in statistical mechanics \cite{izumida,abdoli,iago}. It is to be noted here that
although the probability density function 
(PDF) is known for the linear stochastic process, still, the Fokker-Planck equation is 
required in many cases, as mentioned above. 
Thus, one of the objectives of the present study is to derive the Fokker-Planck equation for the non-Markovian dynamics (of a driven Brownian particle) in a simple way using the PDF.

The theory for the Markovian dynamics is well established, and its application is continued, especially to the newly born subject, stochastic thermodynamics. Now one may cite the famous assertion of van Kampen \cite{van}. He argued that the non-Markovian is the rule, Markovian is an exception \cite{van}.
Study on the non-Markovian dynamics over the last four decades in diverse contexts, such as activated barrier crossing\cite{bdk1,bdk2,bdk3,bdk4,bdk5},
stochastic resonance\cite{sr}, ratchet rectification\cite{ratch}, energy
harvesting\cite{eh} and many others\cite{adel,wang,physha,das,preha,das1,rpreha,oth,oth1,oth2,predas}),
justifies the assertion.  We now note that although the study on the Markovian dynamics of charged
particles started in the early sixties\cite{taylor} of the last century, the present state of knowledge on the non-Markovian dynamics of
the same in the presence of a magnetic field is at an early stage\cite{physha,das,preha,stage,stage1,stage2,das1}.
The reason behind this may be that the theoretical analysis of the non-Markovian
dynamics, in general, is challenging. For further particulars, we refer to the
Refs.\cite{van,mazo}. The probabilistic description of the non-Markovian
dynamics is non-trivial and highly formidable\cite{van,mazo}. 
In the theory of Brownian motion, the Langevin and Fokker-
Planck equations correspond to the coarse-grained model, in which only a few degrees
of freedom related to observables of a Brownian particle(BP) are explicitly retained\cite{mazo}. Assuming the environment of a Brownian particle is a collection of non interecting harmonic oscillators, derivation of the generalized Langevin equation of motion now becomes a common practice in both classical and quantum Brownian motions. Recently it has been derived in Ref.\cite{shraphys} for a Brownian oscillator in the presence of a fluctuating magnetic field. It is to be noted here that starting from basic molecular theory (Liouville equation or the Hamilton's equations of motion), one may obtain the generalized Langevin equation of motion using the projection operator formalism\cite{mazo}. 
Very recently, this formalism has been used in different contexts \cite{pof}. 
One may also obtain the Fokker-Planck equation using the projection operator formalism \cite{mazo}. For the non-Markovian dynamics where the decay of certain molecular correlation functions has an asymptotic slow inverse power decay, the projection operator formalism may not be applicable to derive the Fokker-Planck equation, as a detailed discussion was given in Ref.\cite{mazo}. Then the introduction
of additional variables may thus be essential\cite{van}. Using auxiliary
variables, one may have the continuity equations\cite{oth1,stage1,stage2,fabio,pol,bagexp},
which is an effective Markovian description of a given non-Markovian
design in the extended phase space. In some cases, such as for
a linear system, one may derive the Fokker-Planck equation for the
non-Markovian dynamics without using any additional variable(s)\cite{adel,wang,physha,das,preha,das1,rpreha}.
Using the interaction picture, the Fokker-Planck equations
for the non-Markovian Langevin dynamics with or without isotropic
harmonic force field were first derived in 1976\cite{adel}. Twenty years later, a similar issue with a one-dimensional harmonic oscillator was investigated, introducing a proper characteristic function\cite{wang}.
Applying this approach, the Fokker-Planck equations for the non-Markovian
Langevin dynamics with a magnetic force and a time-dependent force
field were addressed in velocity space and phase space\cite{physha},
respectively. Recently\cite{preha}, it has been extended to
derive the FPE for non-Markovian harmonic oscillator
across a magnetic field and time-dependent force field (TDFF). 
The revised version of this Fokker-Planck equation was given in Ref.\cite{rpreha}. To avoid any confusion, we would mention here that 
there is a typo in the same ((Eq.(7)). 
The quantity, ${\bf {q}}\dot \nabla_{{\bf {x}}}P$ in Eq.(7) must be ${\bf {q}} \dot \nabla_{{\bf {u}}}P$. 

Recently, an alternative solution method has been developed in Ref.\cite{das,das1} in deriving the Fokker-Planck equation for the non-Markovian dynamics. Solving the given Langevin equation of motion, one may obtain the relevant moments as well as the probability distribution function. Knowing the nature of these quantities, a relevant Fokker-Planck equation can be proposed whose coefficients are to be determined applying two simple mathematical notions, (1) Setting of linear algebraic equations with the collection of coefficients of phase space variables and their
appropriate multiples after putting the given distribution function into the proposed Fokker-
Planck equation, and (2) solution of the algebraic equations by the elimination procedure to determine the unknown coefficients. Thus the method seems to be a simple as well
as shortcut one compared to the other solution-based methods\cite{adel,wang}.
It reproduces all the known results. Here it is to be noted that at the present state of knowledge, some of the terms in the FPE for the non-Markovian dynamics in the presence of a magnetic field\cite{preha,rpreha,das1} require special attention. In this context, an essential identification is that the non-Markovian dynamics may induce an electric field from the time-independent magnetic field in the presence of a conservative force field.  Then one may notice that it may modulate diffusion terms. Thus 
one of the key objectives of the present study is to draw special attention to this physical quantity in the field, non-equilibrium statistical mechanics. In addition to this, we demonstrate that how the solution method, proposed by Das et. al.\cite{das,das1} can be used to determine the modulation of the drift term due to an external time dependent deterministic force field in the presence of a frictional meory kernel induced velocity dependent feedback. It is to be noted here that the additional drift term (which is due to the external time-dependent force) in the  proposed Fokker-Planck equation  creates a difficulty in using the solution method \cite{das}. The number of independent relations (a set of linear algebraic equations) among the coefficients which appear in the proposed equation is less than that of the number of relevant unknowns. In this circumstance, we need additional conditions based on the physics of the given system. The solution of the Langevin equation implies that the response function or the susceptibility does not depend on the external force field. Then we consider that the drift terms for the other force fields and the relevant diffusion terms are independent of the external force field. Using this property into the independent relations we determine the modulation of the drift term due to an external deterministic force field in the presence of a frictional memory kernel induced velocity dependent feedback. We show that this technique works even in the presence of both conservative and non-conservative fields, respectively. With four examples, we show that the method works well in this context. This calculation is instructive to determine easily a diffusion term due to an external stochastic force driven Brownian particle in the presence of frictional memory induced velocity dependent feedback, harmonic force and magnetic field. Finally, a reader may find a chronological development of the relevant Fokker-Planck equations such that it will be helpful to understand them.

Before leaving this section, we would like to mention the following point.
Consideration of the time-dependent deterministic force field makes
the present study significantly relevant in the field of stochastic thermodynamics,
which is now at an early stage to consider the non-Markovian dynamics\cite{seifert,predas}.
The connotation of work in ST demands a time-dependent deterministic
force field \cite{seifert,predas,jar}. We mention a few circumstances in the closing remarks. \\

The outlay of the article is as follows. In Sec. II A, we present
an alternative formulation for the derivation of the FPE for the non-Markovian dynamics in the presence of a time-dependent
force field. The Fokker-Planck equation is
derived in the following three consecutive subsections for additional force
from harmonic potential or magnetic field or both of them, respectively.
We conclude the paper in Sec. III.

\section{Memory induced modulation of the effect of a time dependent external force field}

\subsection{Non-Markovian dynamics of a free Brownian
particle in the presence of a time-dependent force field}

The Langevin equation of motion of a free Brownian particle which
is coupled to a non-Markovian thermal bath in the presence of a time-dependent
force field ($\boldsymbol{a}(t)$) can be read\cite{adel,das} as:
\begin{equation}
\dot{\boldsymbol{u}}=-\int_{0}^{t}\gamma(t-\tau)\boldsymbol{u}(\tau)d\tau+\boldsymbol{f}(t)+\boldsymbol{a}(t)\;,\label{eq1}
\end{equation}
\noindent where $\boldsymbol{u}$ is a velocity vector of the particle
with mass, $m=1$. $\boldsymbol{f}(t)$, in the above equation is
a colored Gaussian thermal noise, and it is connected to the time-dependent
damping strength ($\gamma(t)$) through the following fluctuation-dissipation
relation:
\begin{equation}
\langle\boldsymbol{f}(t)\cdot\boldsymbol{f}(t')\rangle=3k_{B}T\gamma(t-t')\;.\label{eq2}
\end{equation}
\noindent
Thus Langevin equation of motion (\ref{eq1}) is a special case that is valid only for stationary noise procees. In other words,
the thermal bath is assumed to be in equilibrium state. It is applicable for other cases also.

We now go through the same steps as in Ref. \cite{das},
with the time dependence of $a$ being the only difference would be needed as clarification. First, we calculate the relevant moments using the solution of the Langevin Eq.(\ref{eq1}). The solution of this equation 
can be read as 
\noindent 
\begin{eqnarray}
\boldsymbol{g}(t) & = & \boldsymbol{u}-\chi(t)\boldsymbol{u}(0)-\int_{0}^{t}\chi(t-\tau)\boldsymbol{a}(\tau)d\tau\nonumber \\
 & = & \boldsymbol{u}-(\boldsymbol{c}+\boldsymbol{p})=\boldsymbol{u}-\boldsymbol{c^{\prime}}\nonumber \\
 & = & \int_{0}^{t}\chi(t-\tau)\boldsymbol{f}(\tau)d\tau\;\;\;.\label{eq7}
\end{eqnarray}
where we have used $\boldsymbol{c}^{\prime}=\boldsymbol{c}+\boldsymbol{p}$
, $\boldsymbol{c}=\chi(t)\boldsymbol{u}(0)$ and $\boldsymbol{p}=\int_{0}^{t}\chi(t-\tau)\boldsymbol{a}(\tau)d\tau$. $\chi(t)$, in the above equation is defined as 
\begin{eqnarray}
\chi(t) & = & \mathcal{L}^{-1}[\tilde{\chi}(z)]\nonumber \\
 & = & \mathcal{L}^{-1}\left[\frac{1}{z+\tilde{\gamma}(z)}\right]\;\;\;.\label{eq3}
\end{eqnarray}
\noindent Here $\mathcal{L}^{-1}$ denotes Laplace inversion and $\tilde{\gamma}(z)$
is the Laplace transform of $\gamma(t)$
\begin{equation}
\tilde{\gamma}(z)=\int_{0}^{t}e^{-zt}{\gamma}(t)dt\;\;\;.\label{eq4}
\end{equation}
\noindent
$chi(t)$ in Eqs.(\ref{eq7}) is known as the response function or the
susceptibility. One may notice here that it is independent of the external force field. Shortly we will find that this notice 
may help to derive the relevant Fokker-Planck equation in a shortcut way.

All the second moments corresponding to the fluctuations (as given
by Eq.(\ref{eq7}) can be represented by the matrix,
$\boldsymbol{\mathcal{A}}(t)$ with $A_{ij}=\langle g_{i}(t)g_{j}(t)\rangle$. Now using Eq.(\ref{eq7}) ,we have 
$\boldsymbol{\mathcal{A}(t)}=A(t)I$, where $I$ corresponds to the relevant identity matrix and 
\begin{eqnarray}
A(t) & = & \left(\frac{3k_{B}T}{m}\right)\left[1-\chi^{2}(t)\right] \;\;\;.\label{eq12}
\end{eqnarray}
Thus for the linear Langevin equation (\ref{eq1}) with the Gaussian
noise, the velocity distribution function can be written as     
\cite{risken,adel,das}

\noindent 
\begin{equation}
P(\boldsymbol{u},\boldsymbol{u(0)};t)=\left[\frac{3}{2\pi|\boldsymbol{\mathcal{A}(t)}|}\right]^{\frac{3}{2}}\exp\left[-\frac{3}{2}\boldsymbol{g}\dag(t)\boldsymbol{\mathcal{A}}^{-1}(t)\boldsymbol{g}(t)\right]\;\;.\label{eq8}
\end{equation}

In the next step, keeping in mind the nature of the distribution function as well as the matrix,
$\boldsymbol{\mathcal{A}}(t)$,  we may propose the relevant Fokker-Planck equation as \cite{das}

\begin{equation}
\frac{\partial P}{\partial{t}}=\boldsymbol{G(t)}\cdot\nabla P+\beta(t)\nabla\cdot\boldsymbol{u}P+H(t)\nabla^{2}P\;\;,\label{eq16}
\end{equation}
\noindent where $\beta(t)$, $\boldsymbol{G(t)}$ and $H(t)$ are
time-dependent quantities to account for the NMD properly.
The first term on the right-hand side of the above equation corresponds
to the drift term due to the deterministic time-dependent force field,
$\boldsymbol{a}(t)$. The remaining second and last terms are usual drift
and diffusion terms, respectively. Now we have to determine the time-dependent
coefficients, making use of the solution of the above equation. 
Using the distribution function (\ref{eq8}) in Eq.(\ref{eq16}), we obtain the following set of linear algebraic equations with the collection of coefficients of phase space variables and their
appropriate multiples 
\begin{equation}
\boldsymbol{G(t)}-\beta(t){\boldsymbol{c^{\prime}}}+\frac{2H(t)\boldsymbol{{c^{\prime}}}}{A(t)}=\frac{\dot{A}(t)\boldsymbol{{c^{\prime}}}}{A(t)}-\dot{\boldsymbol{c^{\prime}}}\;\;\;,\label{n1}
\end{equation}
and 
\begin{equation}
2H(t)-2\beta(t)A(t)=\dot{A}(t)\;\;\;.\label{n2}
\end{equation}
\noindent Using Eq.(\ref{n2}) into Eq.(\ref{n1}) we have
\begin{equation}
\beta(t)\boldsymbol{c^{\prime}}+\boldsymbol{G(t)}=-\dot{\boldsymbol{c^{\prime}}}\;\;\;.\label{n3}
\end{equation}

\noindent 
It is to be noted here that the additional drift term ($\boldsymbol{G(t)}\cdot\nabla P$) in the  proposed Fokker-Planck equation ( which is due to the external time-dependent force) creates a difficulty to use the solution method \cite{das}. Here we find two linear algebraic equations (\ref{n1}-\ref{n2}) with the three unknown coefficients, $\beta(t), G(t)$ and $H(t)$, respectively. In this circumstance, we need additional conditions based on the physics of the given system. The solution of the Langevin equation implies that the response function or the susceptibility does not depend on the external force field. Then we consider that the drift terms for the other force field and the relevant diffusion term are independent of the external force field. In the absence of the time-dependent force, $\boldsymbol{G(t)}=0$\cite{adel,das},
$\boldsymbol{c^{\prime}}=\boldsymbol{c}$ and then the above equation
becomes 
\begin{equation}
\beta(t)=-\frac{\dot{\chi(t)}}{\chi(t)}\;\;\;,\label{eq17}
\end{equation}

\noindent and $H(t)$ is given by

\begin{equation}
H(t)=\frac{1}{6}\chi^{2}(t)\frac{d}{dt}\left[\chi^{-2}(t)A(t)\right]\;\;\;.\label{eq18}
\end{equation}
We note that the coefficient, $\beta(t)$ of the drift
term due to dissipative action, is independent of $\boldsymbol{c}(t)$
, which (is the average velocity) may be arbitrary as it depends on
the initial condition, $\boldsymbol{u}(0)$. In other words, $\beta(t)$
only depends on the response function as expected. Similarly, $A(t)$,
as well as the diffusion coefficient, $H(t)$, are also independent
of $\boldsymbol{c}(t)$ since the dispersion of velocity does not
depend on the average velocity. Thus $\beta(t)$ and $H(t)$ may remain
the same in the presence of time-dependent force field (which changes
only the average velocity) since the response function and the variance,
$A(t)$, do not depend on it. Then from Eq.(\ref{n3}) we have 
\begin{equation}
\boldsymbol{G(t)}=-(\dot{\boldsymbol{p}}+\beta(t)\boldsymbol{p})\;\;\;,\label{eq17a}
\end{equation}

\noindent Thus, the Fokker-Planck equation corresponding to the Langevin
Eq.(\ref{eq1}) is:

\begin{equation}
\frac{\partial P}{\partial{t}}=-\frac{\dot{\chi}(t)}{\chi(t)}\nabla\cdot\boldsymbol{u}P-(\dot{\boldsymbol{p}}+\beta(t)\boldsymbol{p})\cdot\nabla P+\frac{1}{6}\chi^{2}(t)\frac{d}{dt}\left[\chi^{-2}(t)A(t)\right]\nabla^{2}P\;\;\;.\label{eq19}
\end{equation}

\noindent Now, one may check that the distribution function
(\ref{eq8}) is the solution of the above equation. It constitutes
the necessary and sufficient check of the above calculation. The above
equation was derived earlier in Ref.\cite{physha} using the characteristic
function. Then one may easily notice  the simplicity and the shortcuts of the present method. 

For further check,
one can show that the above equation reduces to the standard result\cite{adel,das}
in the absence of the time-dependent force field. Finally, at the Markovian limit,  $\gamma(t-t^{\prime})=2\gamma_0\delta(t-t^{\prime})$ , the response function can be read as,  
$\chi(t)= \exp^{-(\gamma_0(t))}$,  
where $\gamma_0$ is the damping strength. Then we have
$\beta(t)=\gamma_0,
\dot{\boldsymbol{p}}=-\gamma_0\boldsymbol{p}+\boldsymbol{a},  H(t)=\gamma_0 k_BT$ 
and $
\boldsymbol{G(t)}=-\boldsymbol{a}$. Thus  at
the Markovian limit, 
Eq.(\ref{eq19}) reduces to the known result \cite{mazo,risken,chand,nelson,gardiner,kampen},
\begin{equation}
\frac{\partial P(\boldsymbol{u},t)}{\partial{t}}=\gamma_{0}\nabla\cdot\boldsymbol{u}P-\boldsymbol{a}\cdot\nabla P+\gamma_{0}k_{B}T\nabla^{2}P\;\;\;.\label{eqn15}
\end{equation}
Comparing this equation  with that of Eq.(\ref{eq19}), one may uncover in a simple way  how the memory effect (from 
the time-dependent damping strength) can modulate the influence of the time-dependent external force field.

\subsection{Non-Markovian dynamics of time-dependent force driven harmonic oscillator}

For an isotropic harmonic oscillator having a frequency 
$\omega$, Eq.(\ref{eq1}) becomes
\begin{equation}
\ddot{\boldsymbol{x}}+\omega^{2}\boldsymbol{x}(t)+\int_{0}^{t}\gamma(t-\tau)\dot{\boldsymbol{x}}(\tau)d\tau=\boldsymbol{f}(t)+\boldsymbol{a}(t)\;\;\;,\label{eq20}
\end{equation}

\noindent
where $\boldsymbol{x}$ corresponds to the relevant position vector. 

Now following the previous subsection, the probability
density function for the Langevin equation of motion
{[}Eq.(\ref{eq20}){]} can be written as \cite{risken,adel,das}
\begin{equation}
P(\boldsymbol{x},\boldsymbol{u},\boldsymbol{x}(0),\boldsymbol{u}(0);t)=\left(\frac{1}{2\pi}\right)^{3}\left[\frac{1}{|\boldsymbol{\mathcal{A}(t)}|}\right]^{\frac{3}{2}}\exp\left[-\frac{3}{2}\boldsymbol{g}\dag(t)\boldsymbol{\mathcal{A}(t)}^{-1}\boldsymbol{g}(t)\right]\label{eq25}
\end{equation}
\noindent with 
\begin{equation}
\boldsymbol{g}(t)=\begin{bmatrix}\boldsymbol{g_{1}}(t)\\
\boldsymbol{g_{2}}(t)
\end{bmatrix}\;\;,
\end{equation}

\noindent
where
\begin{eqnarray}
\boldsymbol{g_{1}}(t) & = & \boldsymbol{x}(t)-\left[\chi_{\boldsymbol{x}}(t)\boldsymbol{x}_{0}+\chi_{\boldsymbol{u}}(t)\boldsymbol{u}_{0}-\int_{0}^{t}\chi_{\boldsymbol{u}}(\tau)\boldsymbol{a}(t-\tau)d\tau\right]\nonumber \\
 & = & \boldsymbol{x}(t)-\left[\boldsymbol{c_{1}}+\boldsymbol{q_{x}}\right]\nonumber \\
 & = & \boldsymbol{x}(t)-\boldsymbol{{c_{1}}^{\prime}}\nonumber \\
 & = & \int_{0}^{t}\chi_{\boldsymbol{u}}(\tau)\boldsymbol{f}(t-\tau)d\tau \; \;.
 \label{eqa1}
\end{eqnarray}

\noindent From the above equation, the fluctuations in velocity can be
read as 
\begin{eqnarray}
\boldsymbol{g_{2}}(t) & = & \boldsymbol{u}(t)-\left[\dot{\chi}_{\boldsymbol{x}}(t)\boldsymbol{x}_{0}+\dot{\chi}_{\boldsymbol{u}}(t)\boldsymbol{u}_{0}-\int_{0}^{t}\dot{\chi}_{\boldsymbol{u}}(\tau)\boldsymbol{a}(t-\tau)d\tau\right]\nonumber \\
 & = & \boldsymbol{u}(t)-[\dot{\boldsymbol{c_{1}}}+\boldsymbol{p_{x}}]\nonumber \\
 & = & \boldsymbol{u}(t)-\boldsymbol{{c_{2}}^{\prime}}\nonumber \\
 & = & \int_{0}^{t}\dot{\chi}_{\boldsymbol{u}}(\tau)\boldsymbol{f}(t-\tau)d\tau \; \;.
\label{eqa2}
\end{eqnarray}
\noindent
Here we have used $\boldsymbol{c_{1}}=\left[\chi_{\boldsymbol{x}}(t)\boldsymbol{x}_{0}+\chi_{\boldsymbol{u}}(t)\boldsymbol{u}_{0}\right]$,
$\boldsymbol{q_{x}}=\int_{0}^{t}\chi_{\boldsymbol{u}}(\tau)\boldsymbol{a}(t-\tau)d\tau$
and $\boldsymbol{\dot{q_{x}}}=\boldsymbol{p_{x}}$. $\chi_{\boldsymbol{x}}(t)$
and $\chi_{\boldsymbol{u}}(t)$ in Eq.(\ref{eqa1}) are the inverse
Laplace transform of the relations, $\tilde{\chi}_{\boldsymbol{x}}(s)=\frac{\tilde{\gamma}(s)+s}{s^{2}+s\tilde{\gamma}(s)+\omega^{2}}$
and $\tilde{\chi}_{\boldsymbol{u}}(s)=\frac{1}{s^{2}+s\tilde{\gamma}(s)+\omega^{2}}$,
respectively. Then we define the matrix, $\boldsymbol{\mathcal{A}(t)}$
with $A_{ij}=\langle\boldsymbol{g_{i}}(t)\cdot\boldsymbol{g_{j}}(t)\rangle$.

Now the nature of the distribution function leads to propose the following Fokker-Planck equation  as an equivalent description of Eq.(\ref{eq20}),

\begin{eqnarray}
\frac{\partial{P}}{\partial{t}} & = & -\boldsymbol{u}\cdot\nabla_{{\bf {x}}}P+\boldsymbol{G}(t).\nabla_{{\bf {u}}}P+H_{1}(t)\boldsymbol{x}\cdot\nabla_{{\bf {u}}}+H_{2}(t)\nabla_{{\bf {u}}}.{\bf {u}}P\nonumber \\
 & + & H_{3}(t)\nabla_{{\bf {u}}}.\nabla_{{\bf {x}}}P+H_{4}(t){\nabla_{{\bf {u}}}}^{2}P\;\;\;,\label{eq32a}
\end{eqnarray}
\noindent 
where  $\boldsymbol{G}(t), H_{1}(t), H_{2}(t), H_{3}(t)$ and $H_{4}(t)$ are the relevant time-dependent coefficients to account for the NMD properly. The first term on the right-hand side of the above equation is the usual drift term for both Markovian and non-Markovian dynamics, respectively. In other words, here we consider that the time-dependent external force field can not change the definition of velocity. The next term is due to the time-dependent external force field. The drift term corresponding to the harmonic force field appears with the coefficient $H_{1}(t)$. The remaining drift term is due to the dissipative force. We now consider the inclusion of the diffusion terms. It is welknown \cite{adel,wang} that the non Markovian dynamics induces a correlation between the cannonical conjugate pair. This correlation is instructive
to include the diffusion term, $H_{3}(t)\nabla_{{\bf {u}}}.\nabla_{{\bf {x}}}P$.  
Finally, the last term is the usual diffusion term. To avoid any confusion, we would mention here that following Ref.\cite{das,das1,arxiv}, the diffusion terms with other
second derivatives are not considered here.  

We are now in a position to determine the time-dependent coefficients of the above equation. Following the previous section, we have

\begin{equation}
\boldsymbol{G}(t)=-\dot{\boldsymbol{p_{x}}}-H_{2}(t)\boldsymbol{p_{x}}-H_{1}(t)\boldsymbol{q_{x}}\;\;,\label{eq37a}
\end{equation}

\begin{equation}
H_{1}(t)=\tilde{\omega}^{2}(t)\;\;,\label{eq33}
\end{equation}

\begin{equation}
H_{2}(t)=\tilde{\beta}(t)\;\;,\label{eq34}
\end{equation}

\begin{equation}
H_{3}(t)=\frac{k_BT}{\omega^2}\left[\tilde{\omega}^{2}(t)-\omega^2\right]\;\;,\label{eq37}
\end{equation}

\noindent
and

\begin{equation}
H_{4}(t)=\frac{k_BT}{\omega^2}\tilde{\beta}(t)\;\;\;,\label{eq36}
\end{equation}

\noindent
where we have used

\noindent 
\begin{equation}
\tilde{\beta}(t)=-\frac{d\ln\Delta(t)}{dt}\;\;\;,\label{eq38}
\end{equation}

\begin{equation}
\tilde{\omega}^{2}(t)=\Delta^{-1}(t)\left[\ddot{\chi}_{\boldsymbol{u}}(t)\dot{\chi}_{\boldsymbol{x}}(t)-\ddot{\chi}_{\boldsymbol{x}}(t)\dot{\chi}_{\boldsymbol{u}}(t)\right]\label{eq39}
\end{equation}

\noindent and

\begin{equation}
\Delta(t)=\left[\dot{\chi}_{\boldsymbol{u}}(t)\chi_{\boldsymbol{x}}(t)-\dot{\chi}_{\boldsymbol{x}}(t)\chi_{\boldsymbol{u}}(t)\right]\;\;\;.\label{eq40}
\end{equation}

\noindent Thus the required Fokker-Planck equation can be read as
\begin{eqnarray}
\frac{\partial{P}}{\partial{t}} & = & -\boldsymbol{u}\cdot\nabla_{{\bf {x}}}P-\left[\dot{\boldsymbol{p_{x}}}+\tilde{\beta}(t)\boldsymbol{p_{x}}+\tilde{\omega}^{2}(t)\boldsymbol{q_{x}}\right].\nabla_{{\bf {u}}}P
+ \tilde{\omega}^{2}(t)\boldsymbol{x}.\nabla_{{\bf {u}}}P\nonumber \\
& + & \tilde{\beta}(t)\nabla_{{\bf {u}}}\cdot\boldsymbol{u}P 
+ \frac{k_BT}{\omega^2}\left[\tilde{\omega}^{2}(t)-\omega^2\right] \nabla_{{\bf {u}}}.\nabla_{{\bf {x}}}P
+ \frac{k_BT}{\omega^2}\tilde{\beta}(t) {\nabla_{{\bf {u}}}}^{2}P\;\;\;.\label{eq41}
\end{eqnarray}

\noindent 
Now, one can check that the distribution function
(\ref{eq25}) is a solution of the above equation. Thus we arrive to the new Fokker-Planck equation with a shortcut way. Shortly we will compare this equation with its counterpart,i.e., the Markovian case.

For further check,
one may use the condition, $\boldsymbol{a}(t)=0$. Then $\boldsymbol{q_{x}}=\boldsymbol{p_{x}}=0$
as well as $\boldsymbol{G_{1}}(t)=0$ and the above equation reduces
to the standard result\cite{adel,wang,das} 

\begin{eqnarray}
\frac{\partial{P}}{\partial{t}} & = & -\boldsymbol{u}\cdot\nabla_{{\bf {x}}}P
+ \tilde{\omega}^{2}(t)\boldsymbol{x}.\nabla_{{\bf {u}}}P+\tilde{\beta}(t)\nabla_{{\bf {u}}}\cdot\boldsymbol{u}P \nonumber \\
& + & \frac{k_BT}{\omega^2}\left[\tilde{\omega}^{2}(t)-\omega^2\right] \nabla_{{\bf {u}}}.\nabla_{{\bf {x}}}P
+ \frac{k_BT}{\omega^2}\tilde{\beta}(t) {\nabla_{{\bf {u}}}}^{2}P\;\;\;.\label{eq41a}
\end{eqnarray}

At the Markovian limit, Eq.(\ref{eq41}) reduces to 
\begin{eqnarray}
\frac{\partial{P}}{\partial{t}} & = & -\boldsymbol{u}\cdot\nabla_{{\bf {x}}}P+\omega^{2}\boldsymbol{x}.\nabla_{{\bf {u}}}P+\gamma_{0}\nabla_{{\bf {u}}}\cdot\boldsymbol{u}P-\boldsymbol{a(t)}\cdot\nabla_{{\bf {u}}}P+ \gamma_{0}k_{B}T{\nabla_{{\bf {u}}}}^{2}P\;\;\;.\label{eq41b}
\end{eqnarray}
\noindent
In the absence of the time-dependent force ($\boldsymbol{a(t)}=0$)
the above equation becomes the Kramers' equation for the isotropic
harmonic oscillator\cite{kram}. 

Comparing Eq.(\ref{eq41a}) with Eq.(\ref{eq41b}) we find that as a signature of the memory effect induced feedback, the frequency of the harmonic oscillator and the coefficient of the drift term due to dissipative force  become time-dependent for the non-Markovian dynamics. Then one may compare the coefficients of the diffusion terms (due to the momentum diffusion) in the respective Fokker-Planck equations. Here again, it is apparent that the coeffcient for the non-Markovian dyanmics is time-depenent. The diffusion term in Eq.(\ref{eq41a}) with cross derivatives is purely non Markovian origin. Its coefficient easily measures the deviation of the dynamics from the Markovian character.
Finally, comparing Eq.(\ref{eq41}), with Eq.(\ref{eq41b}) we find that how the modulation of the effect of time-dependent external force can be complicated by the conservative and dissipative forces in the presence of the memory effect induced feedback. Thus the present method is very simple to use compared to the other methods to identify the modulation. Then one may expect to apply the present method to identify the modulation of the cyclotron frequency and its response to  the effect of the time-dependent force in the presence of a non-Markovian thermal bath. The following subsection shows that one may achieve this result through a simple as well as shortcut way following subsection (A).

\subsection{Non-Markovian dynamics of a free particle
in the presence of a constant magnetic field and a time-dependent
force field}

\noindent In the presence of a magnetic field, the equation of motion
(\ref{eq1}) becomes \cite{physha}:
\begin{equation}
\dot{\boldsymbol{u}}=\boldsymbol{F}+\boldsymbol{f}(t)+\boldsymbol{a}(t)\;\;\;,\label{eq42}
\end{equation}
where
\begin{equation}
\boldsymbol{F}=-\int_{0}^{t}\gamma(t-\tau)\boldsymbol{u}(\tau)d\tau+\frac{q}{m}(\boldsymbol{u}\times\boldsymbol{B})\;\;\;.\label{eq43}
\end{equation}
\noindent $\boldsymbol{B}=(0,0,B_{z})$, in the above equation is
the applied magnetic field. Thus, the $z$-direction does not experience
the magnetic force. Therefore, we consider that the motion of the Brownian
particle is confined in $x-y$ plane. The related equations
of motion in terms of components of velocity are:
\begin{equation}
\dot{u_{x}}=-\int_{0}^{t}\gamma(t-\tau)u_{x}(\tau)d\tau+\Omega u_{y}+f_{x}(t)+a_{x}(t)\label{eq46}
\end{equation}
\noindent and
\begin{equation}
\dot{u_{y}}=-\int_{0}^{t}\gamma(t-\tau)u_{y}(\tau)d\tau-\Omega u_{x}+f_{y}(t)+a_{y}(t)\;\;\;\label{eq47}
\end{equation}
where we have used $\Omega=\frac{qB_{z}}{m}$. $a_{x}(t)$ and $a_{y}(t)$
are the relevant components of the time-dependent force, $\boldsymbol{a}(t)$.

Now the velocity distribution
function for the Langevin Eqs.(\ref{eq46}-\ref{eq47}) can be written as 
\begin{equation}
P(u_{x},u_{y},u_{x}(0),u_{y}(0);t)=\left(\frac{1}{2\pi}\right)\left[\frac{1}{|\boldsymbol{\mathcal{A}(t)}|}\right]^{\frac{1}{2}}\exp\left[-\frac{1}{2}g\dag(t)\boldsymbol{\mathcal{A}}^{-1}(t)g(t)\right]\label{eq50}
\end{equation}
\noindent with 
\begin{equation}
g(t)=\begin{bmatrix}g_{1}(t)\\
g_{2}(t)
\end{bmatrix}\;\;.\label{eq51}
\end{equation}

\noindent
where

\begin{eqnarray}
g_{1}(t) & = & u_{x}(t)-\left[\left(\chi_{1}(t)u_{x}(0)+\chi_{2}(t)u_{y}(0)\right)+\left(\int_{0}^{t}\chi_{1}(\tau)a_{x}(t-\tau)d\tau+\int_{0}^{t}\chi_{2}(\tau)a_{y}(t-\tau)d\tau\right)\right]\nonumber \\
 & = & u_{x}(t)-\left[c_{1}+p_{x}\right]\nonumber \\
 & = & u_{x}(t)-{c_{1}}^{\prime}\nonumber \\
 & = & \int_{0}^{t}\chi_{1}(\tau)f_{x}(t-\tau)d\tau+\int_{0}^{t}\chi_{2}(\tau)f_{y}(t-\tau)d\tau\label{eq48}
\end{eqnarray}

\noindent and

\begin{eqnarray}
g_{2}(t) & = & u_{y}(t)-\left[\left(\chi_{1}(t)u_{y}(0)-\chi_{2}(t)u_{x}(0)\right)+\left(\int_{0}^{t}\chi_{1}(\tau)a_{y}(t-\tau)d\tau-\int_{0}^{t}\chi_{2}(\tau)a_{x}(t-\tau)d\tau\right)\right]\nonumber \\
 & = & u_{y}(t)-\left[c_{2}+p_{y}\right]\nonumber \\
 & = & u_{x}(t)-{c_{2}}^{\prime}\nonumber \\
 & = & \int_{0}^{t}\chi_{1}(\tau)f_{y}(t-\tau)d\tau-\int_{0}^{t}\chi_{2}(\tau)f_{x}(t-\tau)d\tau\;\;\;.\label{eq49}
\end{eqnarray}

\noindent Here we have used $c_{1}=\chi_{1}u_{x}(0)+\chi_{2}u_{y}(0)$
, $c_{2}=\chi_{1}u_{y}(0)-\chi_{2}u_{x}(0)$, $p_{x}=\int_{0}^{t}\chi_{1}(\tau)a_{x}(t-\tau)d\tau+\int_{0}^{t}\chi_{2}(\tau)a_{y}(t-\tau)d\tau$
and $p_{y}=\int_{0}^{t}\chi_{1}(\tau)a_{y}(t-\tau)d\tau-\int_{0}^{t}\chi_{2}(\tau)a_{x}(t-\tau)d\tau$.
The response functions, $\chi_{1}(t)$ and $\chi_{2}(t)$, which appear
in the above equations are inverse Laplace transform of the following
relations

\begin{equation}
\tilde{\chi}_{1}(s)=\frac{\left[s+\tilde{\gamma}(s)\right]}{\left[s+\tilde{\gamma}(s)\right]^{2}+\Omega^{2}},\label{eq49a}
\end{equation}

\noindent and

\begin{equation}
\tilde{\chi}_{2}(s)=\frac{\Omega}{\left[s+\tilde{\gamma}(s)\right]^{2}+\Omega^{2}},\label{eq49b}
\end{equation}

\noindent respectively. Then we define the matrix, $\boldsymbol{\mathcal{A}(t)}$
with $A_{ij}=\langle g_{i}(t)g_{j}(t)\rangle$.

We are now in a position to propose the
Fokker-Planck equation for the non-Markovian dynamics with Eqs.(\ref{eq46}-\ref{eq47})\cite{das} as
\begin{eqnarray}
 \frac{\partial{P}}{\partial{t}} & = & \beta_{1}\left[\nabla_{{\bf {u}}}.{\bf {u}}P\right]+\beta_{2}\left[{\bf {u}}\times\nabla_{{\bf {u}}}P\right]_{z}+G_{1}(t)\frac{\partial P}{\partial u_{x}}+G_{2}(t)\frac{\partial P}{\partial u_{y}}+H(t){\nabla_{{\bf {u}}}}^2 P\;\;\;,\label{eq59}
 \end{eqnarray}
\noindent where $\beta_{1}(t)$, $\beta_{2}(t)$, $H(t)$ ,$G_{1}(t)$
and $G_{2}(t)$ are the relevant time-dependent quantities. 

Following  Sec. IIA, we determine these coefficients as 
\begin{equation}
  \beta_{1}(t)=-\frac{(\chi_{1}\dot{\chi}_{1}+\chi_{2}\dot{\chi}_{2})}{(\chi_{1}^{2}+\chi_{2}^{2})}\;\;\;,\label{eq60}
  \end{equation}
\begin{equation}
\beta_{2}(t)=-\frac{(\chi_{2}\dot{\chi_{1}}-\chi_{1}\dot{\chi}_{2})}{(\chi_{1}^{2}+\chi_{2}^{2})}\;\;\;,\label{eq61}
\end{equation}
\begin{equation}
H(t)=\frac{(\chi_{1}^{2}+\chi_{2}^{2})}{2}\frac{d}{dt}\left[\frac{A(t)}{(\chi_{1}^{2}+\chi_{2}^{2})}\right]\;\;,\label{eq62}
\end{equation}

\begin{equation}
G_{1}(t)=-\dot{p_{x}}-\beta_{1}(t)p_{x}+\beta_{2}(t)p_{y}\;\;\;,\label{eq62a}
\end{equation}
\noindent and 
\begin{equation}
G_{2}(t)=-\dot{p_{y}}-\beta_{1}(t)p_{y}-\beta_{2}(t)p_{x}\;\;\;.\label{eq62c}
\end{equation}
\noindent Making use of Eqs.(\ref{eq60}-\ref{eq62c}) in Eq.(\ref{eq59}),
we have
 \begin{eqnarray}
 \frac{\partial{P}}{\partial{t}} & = & \beta_{1}(t)\nabla_{{\bf {u}}}.{\bf {u}}P-\left[\bf {\dot{p}}+\beta_{1}(t){\bf {p}}\right] .\nabla_{{\bf {u}}}P - \beta_{2}(t)\left[{\bf {p}}\times\nabla_{{\bf {u}}}P\right]_{z} \nonumber\\
  & + & \beta_{2}(t)\left[{\bf {u}}\times\nabla_{{\bf {u}}}P\right]_{z}+H(t){\nabla_{{\bf {u}}}}^{2}P\;\;\;.\label{eq63}
  \end{eqnarray}

\noindent where ${\bf {p}}=\left(p_{x},p_{y},0\right)$. This is the required Fokker-Planck equation.  Now one can check that the distribution
function (\ref{eq50}) is a solution to the above equation. It signifies
an essential appraisal of the present calculation. Another check is
that the above equation exactly corresponds to the one which was derived in Ref.\cite{physha} using the characteristic function. Now comparing
this subsection with Sec.2 in Ref.\cite{physha} one may be sure about the simplicity and the shortcuts of the present method. The plainness may be noticeable in the additional harmonic force field, which
we will demonstrate in the next subsection. 

For further check, one can show easily that in the absence of time-dependent
force, $\boldsymbol{a(t)}=0$ and the above equation reduces to the
following known result\cite{physha,das}, 

\begin{equation}
 \begin{array}{rcl}
 \dfrac{\partial{P}}{\partial{t}} & = & \beta_{1}(t)\left[\nabla_{{\bf {u}}}.{\bf {u}}P\right]\end{array}+\beta_{2}(t)\left[{\bf {u}}\times\nabla_{{\bf {u}}}P\right]_{z}+H(t){\nabla_{{\bf {u}}}}^{2}P\;\;\;,\label{eq63a}
\end{equation}

since $p_{x}=p_{y}=0$ as well as $G_{1}=G_{2}=0$.

Finally, at the Markovian limit, the Fokker-Planck
equation (\ref{eq63}) reduces to 

\begin{equation}
 \dfrac{\partial{P}}{\partial{t}}  =  \gamma_{0}\left[\nabla_{{\bf {u}}}.{\bf {u}}P\right]-{\bf {a}.}\nabla_{{\bf {u}}}P
 +\Omega\left[{\bf {u}}\times\nabla_{{\bf {u}}}P\right]_{z}+\gamma_{0}k_{B}T{\nabla_{{\bf {u}}}}^{2}P
 \label{eq63n}
 \end{equation}

\noindent In the absence of a time-dependent force field, $\boldsymbol{a}(t)=0$,
the above equation is the same as that obtained for the Markovian
Brownian motion is described by a charged particle across a magnetic
field \cite{ferrari,czopnik}. Thus  
the Fokker-Planck equation (\ref{eq63}) satisfies all possible limiting
situations. However, it is to be noted here that comparing the above equation with that of Eqs.(\ref{eq63}-\ref{eq63a}), one may find how the cyclotron frequency, as well as the effect of the time-dependent external force field, is modulated in the presence of a non-Markovian thermal bath. The modulations are quite similar  to the case, the external time-dependent force driven harmonic oscillator. Here the cyclotron frequency plays a similar role as that of the frequency of the harmonic oscillator.
In the following subsection, we will investigate the relevant modulations in the presence of both conservative and non-conservative force fields, respectively.

\subsection{Non-Markovian dynamics of a Brownian oscillator in the presence of a magnetic field and a time-dependent external force}

In the presence of a two-dimensional harmonic potential energy field,
the Langevin equations of motion (\ref{eq46}-\ref{eq47}) become
\cite{preha} 
\begin{equation}
\dot{u}_{x}=-\omega^{2}x-\int_{0}^{t}\gamma(t-\tau)u_{x}(\tau)d\tau+\Omega u_{y}+f_{x}(t)+a_{x}(t)\label{leq1}
\end{equation}
\noindent and
\begin{equation}
\dot{u}_{y}=-\omega^{2}y-\int_{0}^{t}\gamma(t-\tau)u_{y}(\tau)d\tau-\Omega u_{x}+f_{y}(t)+a_{y}(t)\;\;\;.\label{leq2}
\end{equation}
\noindent where $\omega$ is the frequency of the harmonic oscillator.
Since the Langevin equations (\ref{leq1}-\ref{leq2}) of motion correspond
to the Gaussian noise-driven linear system, then the phase space distribution
function is a Gaussian one \cite{risken}. Using the matrix, $\boldsymbol{\mathcal{A}}^{\prime}$
(which is defined in Appendix A) and its inverse, the phase space distribution function can be written as:
\begin{equation}
P(x,x(0);y,y(0);u_{x},u_{x}(0);u_{y},u_{y}(0);t)=(2\pi)^{-2}(A_{1}A_{2}-A_{3}^{2}-A_{4}^{2})^{-1}\exp\left[-\frac{1}{2}g\dag(t){\boldsymbol{\mathcal{A}}^{\prime}}^{-1}(t)g(t)\right]\label{eq76}
\end{equation}
with 
\begin{equation}
g(t)=\begin{bmatrix}g_{1}(t)\\
g_{2}(t)\\
g_{3}(t)\\
g_{4}(t)
\end{bmatrix}.\label{eq77}
\end{equation}
\noindent
$g_1(t)$, $g_2(t)$, $g_3(t)$ and $g_4(t)$ are defined in Appendix A.
Then, considering the previous subsections, one may propose the following
Fokker-Planck equation, 

\begin{eqnarray}
\frac{\partial P}{\partial t} & = & -\frac{\partial u_{x}P}{\partial x}-\frac{\partial u_{y}P}{\partial y} - G_{1}(t)\frac{\partial P}{\partial u_{x}}-G_{2}(t)\frac{\partial P}{\partial u_{y}}+H_{1}(t)\left[x\frac{\partial P}{\partial u_{x}}+y\frac{\partial P}{\partial u_{y}}\right]\nonumber \\
& + & H_{2}(t)\left[\frac{\partial u_{x}P}{\partial u_{x}}+\frac{\partial u_{y}P}{\partial u_{y}}\right]
-  H_{3}(t)\left[\frac{\partial u_{y}P}{\partial u_{x}}-\frac{\partial u_{x}P}{\partial u_{y}}\right]
-  H_{4}(t)\left[x\frac{\partial P}{\partial u_{y}}-y\frac{\partial P}{\partial u_{x}}\right]\nonumber \\
& + & H_{5}(t)\left[\frac{\partial}{\partial x}\frac{\partial P}{\partial u_{y}}-\frac{\partial}{\partial y}\frac{\partial P}{\partial u_{x}}\right]+ H_{6}(t)\left[\frac{\partial}{\partial x}\frac{\partial P}{\partial u_{x}}+\frac{\partial}{\partial y}\frac{\partial P}{\partial u_{y}}\right]+H_{7}(t)\left[\frac{\partial^{2}P}{\partial u_{x}^{2}}+\frac{\partial^{2}P}{\partial u_{y}^{2}}\right]\nonumber \\
\label{eq84}
\end{eqnarray}

where $G_{1}(t),G_{2}(t),H_{1}(t),H_{2}(t),H_{3}(t),H_{4}(t),H_{5}(t),H_{6}(t)$ and $H_{7}(t)$
are relevant time-dependent quantities to account for the non-Markovian
dynamics properly. The terms with $G_{1}(t)$ and $G_{2}(t)$ appear to consider the modulation of the effect of the time dependent force 
by the harmonic, the magnetic and the dissipative force fields in the presence of the memory effect induced feedback. Similarly, $H_{1}(t)$ ( $H_{3}(t)$) reperesnts the modulation of the force constant of the Harmonic oscillator (the cyclotron frequency) by the magnetic field (harmonic force field) and the frictional memory kernel. Then we would mention that the term with $H_{2}(t)$ appears to consider the effect of the dissipative force in the presence of magnetic field and harmonic force. The remaining drift term with $H_{4}(t)$ is instructive from the last subsection by virtue of the memory induced time dependent cyclotron frequecny. Thus the Brownian particle experiences an effective time dependent magnetic field from a time independent one in the presence of a non-Markovian thermal bath. Then it is expected to consider a drift term with $H_{4}(t)$ (due to an accociated induced electric field) in the Fokker-Planck equation in phase space. We now address the inclusion
of the diffusion terms. The matrix, $\boldsymbol{\mathcal{A}}^{\prime}$ with the second moments includes the non-Markovian dynamics induced  special kind of cross-correlations, $<xu_y>$ and $<yu_x>$ in the presence of magnetic field. These correlations are  instructive to consider a diffusion term with $H_{5}(t)$ which contains cross derivatives.  Finally, $H_{6}(t)$ and $H_{7}(t)$ are the usual diffusion terms for the non-Markovian dynamics in the phase space as suggested by $\boldsymbol{\mathcal{A}}^{\prime}$. To avoid any confusion, we would mention 
that the diffusion terms with other possible cross derivatives are
not considered since the cross-correlation of the fluctuations is
zero for the individual case. However, following the previous subsections, we obtain the required Fokker-Planck equation as

\begin{eqnarray}
\frac{\partial P}{\partial t} & = & -{\bf {u}}.\nabla_{{\bf {x}}}P
-\left[\left[{\bf {\dot{p}}}+H_{2}(t){\bf {p}}+H_{1}(t){\bf {q}}\right].{\nabla_{{\bf {u}}}}P+H_{3}(t)\left[{\bf {p}}\times\nabla_{{\bf {u}}}P\right]_{z}-H_{4}(t)\left[{\bf {q}}\times\nabla_{{\bf {u}}}P\right]_{z}\right]\nonumber \\
 & + & H_{1}(t){\bf {x}}.\nabla_{{\bf {u}}}P+H_{2}(t)\nabla_{{\bf {u}}}.{\bf {u}}P+H_{3}(t)\left[{\bf {u}}\times\nabla_{{\bf {u}}}P\right]_{z}-H_{4}(t)\left[{\bf {x}}\times\nabla_{{\bf {u}}}P\right]_{z}\nonumber \\
 & + & H_{5}(t)\left[\nabla_{{\bf {x}}}\times\nabla_{{\bf {u}}}P\right]_{z}+H_{6}(t)\nabla_{{\bf {u}}}.\nabla_{{\bf {x}}}P+H_{7}(t){\nabla_{{\bf {u}}}}^{2}P\;\;.\label{eq84-1}
\end{eqnarray}

\noindent
Here $\bf {x}$ and $\bf {u}$ are relevant position and velocity vectors, respectively. The time-dependent coefficients in the above equations are defined in Appendix B. Using the definition of the coefficients, one may check that
the distribution function ({\ref{eq76}}) is a solution of the above
Fokker-Planck equation.

For further checking of the present calculation, we consider the condition with $\boldsymbol{a}(t)=0$.
Then one can easily  show  that $\bf{\dot{p}}=\bf {p}=\bf {q}=0$ and the
above Fokker-Planck equation reduces to the following known result\cite{das1}

\begin{eqnarray}
\frac{\partial P}{\partial t} & = &  -{\bf {u}}.\nabla_{{\bf {x}}}P+H_{1}(t){\bf {x}}.\nabla_{{\bf {u}}}P+H_{2}(t)\nabla_{{\bf {u}}}.{\bf {u}}P+H_{3}(t)\left[{\bf {u}}\times\nabla_{{\bf {u}}}P\right]_{z}\nonumber \\
 & - & H_{4}(t)\left[{\bf {x}}\times\nabla_{{\bf {u}}}P\right]_{z}+H_{5}(t)\left[\nabla_{{\bf {x}}}\times\nabla_{{\bf {u}}}P\right]_{z}+H_{6}(t)\nabla_{{\bf {u}}}.\nabla_{{\bf {x}}}P+H_{7}(t){\nabla_{{\bf {u}}}}^{2}P\;\;\;.\label{eq84a}
\end{eqnarray}
\noindent
At the same time, Eq.(\ref{eq84-1}) also reduces to all the other known results at the specific limits. Thus the accuracy of the calculation is confirmed. 

Before leaving this subsection, we would mention the following points.
The Fokker-Planck equation corresponding to the equations of motion (\ref{leq1}-\ref{leq2}) was derived in Ref.\cite{preha} using the relevant characteristic function. This equation was not a correct one as reported in \cite{das1}. Then its revised version (Eq.(7) in \cite{rpreha}) was published very recently. There is a typo in Eq.(7), as mentioned in the introduction. However, some comments in \cite{rpreha} regarding the present method seem to be misleading. Then we put the following discussion.

\noindent
(i) To apply the present method, one may not need
prior knowledge based on the characteristic function. For example, the proposed Fokker-Planck equations in Ref.\cite{das1} contain
extra terms compared to the required Fokker-Planck equation.
Applying two simple mathematical notions [(1) Setting of linear algebraic equations with the collection of coefficients of phase space variables and their appropriate multiples after putting the given distribution function into the proposed Fokker-Planck equation, and  (2)
solution of the algebraic equations by the elimination procedure.], one may anchor at the required FPE. It leads to remove the
inconsistencies with two drift terms and a diffusion term in the Fokker-Planck equation (61) in Ref.\cite{preha}.
Thus the method seems to be a straightforward as well as shortcut one compared to the other solution-based methods\cite{adel,wang}. It may be instructive to derive the Fokker-Planck equation in a shortcut way for complex systems like the magnetic field in an arbitrary direction.

\noindent
(ii)  The present method may be helpful for a more complex case like a multidimensional system with  
relevant position-dependent couplings. In that case, the drift term from the harmonic force field has to be split. Thus the present method may be pertinent for all possible Gaussian noise-driven linear systems for which the distribution function is known. The relevant matrix (with the second moments corresponding to the component of fluctuations) in the distribution function may be diagonal or not. For example, in subsections II A and II C, the matrix is diagonal, and all the off-diagonal elements may not be zero in subsections II B and D, respectively.

\noindent
(iii)
Then we address the issue regarding the shorcut way  of the method. Comparing between Sec.IIC of the present manuscript and Sec.2 plus alied Appendix B in \cite{physha},
one may be sure about the shorcutness of the present calculation. To avoid any confusion, we note that even some algebra was
not explicit in this reference. Specifically, the transformation of Eq.(B.10) into  Eq.(B.22) with the use of Eqs.(B.9) and (B.17-B.18) requires
a lengthy algebra. It is to be noted here that in the Sec.IIC, three algebraic relations among five unknown, $\beta_{1}(t),  \beta_{2}(t), G_1(t), G_2(t)$ and $H(t)$ are implicit. One may obtain these relations upon substitution of the distribution function(\ref{eq50}) in Eq.(\ref{eq59}). $\beta_{1}(t), \beta_{2}(t)$ and $H(t)$ are independent of the external time-dependent force field. Using this property, one can easily determine these quantities as reported in \cite{das}. Then determination of $G_1(t)$ and $G_2(t)$ is very simple. Thus the relevant algebra is almost the same  with or without the external time-dependent force field. Thus the present method is too shortcut (in adition to its simplicity) compared to other method. Furthermore, one may compare Ref.\cite{arxiv} with Sec.III  allied with appendix A in \cite{preha}  to avoid any confusion regarding the present context. Again we note that a very long and careful algebra was implicit, as mentioned in the reference \cite{preha}. It is to be noted here that the algebraic equations in the appendix in Ref.\cite{arxiv}  were solved by a simple inspection. This leads to determine $G_1(t)$ and $G_2(t)$ in Eq.(\ref{eq84}) in Sec.IID with a bit of additional effort to derive the relations among the coefficients in this equation. Thus there should be no doubt about the shortcut way of the present method.

\subsection{Memory induced electric field from time-independent magnetic field} 

    There is no particular emphasis in all the relevant references \cite{das,preha,das1,rpreha} to interpret the physical significance of the additional terms (in the required Fokker-Planck equations) due to the applied magnetic field. This subsection includes an approach to this issue. \\

In the presence of a time-dependent magnetic field ($\boldsymbol{B_t}=(0,0,B_t)$), the Langevin equations of motion (\ref{leq1}-\ref{leq2}) can be written as \cite{shraphys} 
\begin{equation}
\dot{u}_{x}=-\omega^{2}x-\int_{0}^{t}\gamma(t-\tau)u_{x}(\tau)d\tau+\Omega_t u_{y}+\frac{\dot{B_t} y}{2}+f_{x}(t)+a_{x}(t)\label{leqel1}
\end{equation}

\noindent and
\begin{equation}
\dot{u}_{y}=-\omega^{2}y-\int_{0}^{t}\gamma(t-\tau)u_{y}(\tau)d\tau-\Omega_t u_{x}-\frac{\dot{B_t} x}{2}+f_{y}(t)+a_{y}(t)\label{leqel2} \; \;,
\end{equation}

\noindent
where $\Omega_t=\frac{qB_t}{m}$. At the Markovian limit, the overhead equations of motion evolves as:
\begin{equation}
\dot{u}_{x}=-\omega^{2}x-\gamma_0u_{x}+\Omega_t u_{y}+\frac{\dot{B_t} y}{2}+f_{x}(t)+a_{x}(t)\label{leqel3}
\end{equation}
\noindent and
\begin{equation}
\dot{u}_{y}=-\omega^{2}y-\gamma_0 u_{y}-\Omega_t u_{x}-\frac{\dot{B_t} x}{2}+f_{y}(t)+a_{y}(t)\label{leqel4} \; \;.
\end{equation}

Now making use of the respective Taylor series into the both sides of the Smoluchowski integral equaion (which is the Chapman-Kolmogorov equation for a Markov process \cite{mazo}), essentially due to Einstein \cite{ein},
corresponding to the above equations of motion, we obtain the following Fokker-Planck equation, 
\begin{eqnarray}
\frac{\partial P}{\partial t} & = & -{\bf {u}}.\nabla_{{\bf {x}}}P+\gamma_0 \nabla_{{\bf {u}}}.{\bf {u}}P -\bf {a}.\nabla_{{\bf {u}}}P +\omega^{2}{\bf {x}}.\nabla_{{\bf {u}}}P+\Omega_t\left[{\bf {u}}\times\nabla_{{\bf {u}}}P\right]_{z} \nonumber\\
& + & \frac{\dot{B_t}}{2}\left[{\bf {x}}\times\nabla_{{\bf {u}}}P\right]_{z}+\gamma_0 k_BT{\nabla_{{\bf {u}}}}^{2}P \;\;.\label{eqel84-1}
\end{eqnarray}

\noindent
Adding the noise term to the relevant Liouville equation and then doing proper averaging of the equation\cite{zwang}, one may easily arive at  this equation. However, in the absence of the external time-dependent force field, the above equation evolves to
\begin{eqnarray}
\frac{\partial P}{\partial t} & = & -{\bf {u}}.\nabla_{{\bf {x}}}P+\gamma_0 \nabla_{{\bf {u}}}.{\bf {u}}P +\omega^{2}{\bf {x}}.\nabla_{{\bf {u}}}P+\Omega_t\left[{\bf {u}}\times\nabla_{{\bf {u}}}P\right]_{z}\nonumber \\
 & + & \frac{\dot{B_t} }{2}\left[{\bf {x}}\times\nabla_{{\bf {u}}}P\right]_{z}+\gamma_0 k_BT{\nabla_{{\bf {u}}}}^{2}P\;\;.\label{eqel84-2}
\end{eqnarray}

Similarly, the Fokker-Planck equation at the Markovian limit for the Langevin equations (\ref{leq1}-\ref{leq2}) of motion can be written as

\begin{eqnarray}
\frac{\partial P}{\partial t} & = & -{\bf {u}}.\nabla_{{\bf {x}}}P-\bf {a}.\nabla_{{\bf {u}}}P +\omega^{2}{\bf {x}}.\nabla_{{\bf {u}}}P+\gamma_0 \nabla_{{\bf {u}}}.{\bf {u}}P+\Omega\left[{\bf {u}}\times\nabla_{{\bf {u}}}P\right]_{z} +\gamma_0 k_BT{\nabla_{{\bf {u}}}}^{2}P \;\;.\label{eqel84-3}
\end{eqnarray}

\noindent
Now one may discourse on the following points in order. First, comparing Eq.(\ref{eq84a}) with Eq.(\ref{eqel84-2}), we find that the
coefficient of $\left[{\bf {u}}\times\nabla_{{\bf {u}}}P\right]_{z}$ is time-dependent in both equations. Then it is apparent in Eq.(\ref{eq84a}) that the time-independent magnetic field behaves effectively as a time-dependent one in the presence  of a non-Markovian thermal bath. 

Then one may expect the frictional memory kernel induced electric field from the time-independent magnetic field in the phase space description. In this occurrence, we would look at Eq.(\ref{eqel84-2}). The term, $\frac{\dot{B_t} }{2}\left[{\bf {x}}\times\nabla_{{\bf {u}}}P\right]_{z}$ in this equation is due to the induced electric field from the time-dependent magnetic field. Then one may identify the term,
with the coefficient $H_4\left[{\bf {x}}\times\nabla_{{\bf {u}}}P\right]_{z}$ in Eq. (\ref{eq84a}) as the memory-induced electric field from the time-independent magnetic field. Here the conservative force field from the harmonic potential has an important role since, in the absence, the induced electric field may not emerge as justified in Eq.(23) in Ref.\cite{arxiv}. Second, analogizing among the Fokker-Planck Eqs.(\ref{eq63}, \ref{eq84-1},\ref{eqel84-1}), one may interpret that the drift term, $H_{4}(t) \left[{\bf {q}}\times\nabla_{{\bf {u}}}P\right]_{z}$  in Eq.(\ref{eq84-1}) is due to the modulation of the effect of the time-dependent external force field by the memory induced electric field. Thus, the second term in the right hand side of this equation exhibits how the exffect can be modulated by frictional memory kernel, conservative force, constant magnetic field and induced electric field.
Finally,  the terms with $H_{5}(t)$ and $H_{7}(t)$ in Eq.(\ref{eq84-1}) imply that how the diffusion can be modulated by the electric field due to the non Markovian dyamics. These are important benefits of the derivation of the Fokker-Planck equation. The induced electric field and related effects may be overlooked from the Langevin equations of motion. Even the existence of these quantities may be identified with the help of the distribution function. Still, the lengthy expression of the PFE equation implies that the identification can not  determine the effects' contribution to the probability flux quantitatively. According to the famous assertion of van Kampen \cite{van},  the identification of the induced electric field demands to re-investigate the barrier crossing dynamics \cite{abdoli,alen,alen1,shra,shraphys,barri} and others \cite{baurapre,jayann}, where it was considered that a charged Brownian particle is coupled to a Markovian thermal bath in the presence of  magnetic field and conservative force.

\section{Conclusion}

 In the presnt study we demonstrate that how the solution method, proposed by Das et. al.\cite{das,das1} can be used to determine the modulation of the drift term due to an external time dependent deterministic force field in the presence of a frictional meory kernel induced velocity dependent feedback. It is to be noted here that the additional drift term (which is due to the external time-dependent force) in the  proposed Fokker-Planck equation  creates a difficulty in using the solution method \cite{das}. The number of independent relations (a set of linear algebraic equations) among the coefficients which appear in the proposed equation is less than that of the number of relevant unknowns. In this circumstance, we need additional conditions based on the physics of the given system. The solution of the Langevin equation implies that the response function or the susceptibility does not depend on the external force field. Then we consider that the drift terms for the other force fields and the relevant diffusion terms are independent of the external force field. Using this property into the independent relations we determine the modulation of the drift term due to an external deterministic force field in the presence of a frictional memory kernel induced velocity dependent feedback. We show that this technique works even in the presence of both conservative and non-conservative fields, respectively. With four examples, we show that the method works well in this context. This calculation is instructive to determine easily a diffusion term due to an external stochastic force driven Brownian particle in the presence of frictional memory induced velocity dependent feedback, harmonic force and magnetic field. Finally, one may find a chronological development of the relevant Fokker-Planck equations such that it will be helpful to understand them.

It is to be noted that at the present state of knowledge, some of the terms in the FPE for the non-Markovian dynamics in the presence of a magnetic field\cite{preha,rpreha,das1} require special attention. In this context, an essential identification is that the non-Markovian dynamics may induce an electric field from the time-independent magnetic field in the presence of a conservative force field. Thus one may notice that how it may modulate diffusion terms and the effect from a time-dependent external force field.

Before we terminate, the following points merit due concentration. First, one may apply the present method to consider where the magnetic field may be in an arbitrary direction. It seems to be an open problem. In this context, Sec. IID may be instructive in deriving the Fokker-Planck equation with the least effort. Second, the magnetic field induced additional drift and diffusion terms in the company of a non-Markovian thermal bath may be pretty interesting in many contexts. Then the consideration of the time-dependent deterministic force field
makes the present study very relevant in the field of stochastic thermodynamics,
which is now at an early stage of  considering the non-Markovian dynamics\cite{seifert,predas}.
The definition of work in ST requires a time-dependent deterministic
force field to drive the system from a given equilibrium state\cite{seifert,predas,jar}. Then the present study may
be applicable to investigate many aspects of stochastic thermodynamics,
such as (i) how the time asymmetry inherent to irreversible processes
depends on the non-Markovian dynamics? (ii) the effect of non-Markovian
dynamics on the relative entropy and dispersed works, (iii) work
produced from a single reservoir, (iv) characterization of a non-equilibrium steady state in the presence of a non-Markovian thermal
bath etc.

Third, in recent technology, the study of the ion-conducting
electrolytic materials is a significant area in physics and chemistry\cite{scros,angel}.
The materials have potential applications in a diverse range of all-solid-state
devices, such as rechargeable lithium batteries, flexible electrochromic
displays and smart windows\cite{scros}. The properties of the electrolytes
are tuned by varying chemical composition to a large extent and hence
are adapted to specific needs \cite{angel}. High ionic conductivity
is needed for optimizing the glassy electrolytes in various applications.
Then it would be fascinating to tune the ionic conductivity
according to a specific need by a physical method. In this context,
very recent studies \cite{stage2,alen,alen1,shra,shraphys} shows
that an applied magnetic field can tune the conductivity of electrolytic material. To adjust the conductivity of ions in the
solid electrolytes, the combination of both magnetic field and time-dependent
electric field may be an important choice. Thus the present study
may find crucial applications in studying the barrier crossing dynamics \cite{stage2,alen}.

Finally, based on the recent developments on the noise-induced transition in a fluctuating magnetic field-driven harmonic oscillator, one may anticipate that the memory-induced electric field may render the transition even in an additive noise-driven linear system. The relevant issues are in progress.

\section{Acknowledgment}

\label{s6}  M. Biswas is
happy to acknowledge JRF from the Council of Scientific and Industrial
Research, Government of India. D. Mondal thanks SERB (Project No. ECR/2018/002830/CS), Department of Science and Technology, Government of India, for financial support and IIT Tirupati for the new faculty seed grant.


\appendix

\section{Solution of Eqs. (\ref{leq1}-\ref{leq2}) and related quantities}

Using the Laplace transformation the solution of Eqs. (\ref{leq1}-\ref{leq2})
can be written as

\begin{eqnarray}
g_{1}(t) & = & x(t)-\left(A(t)x(0)-B(t)y(0)+C(t)u_{x}(0)+D(t)u_{y}(0)\right)\nonumber \\
 & - & \left(\int_{0}^{t}H_{0}(t-\tau)a_{x}(\tau)d\tau-\Omega^{2}\int_{0}^{t}H_{0}^{\prime}(t-\tau)a_{x}(\tau)d\tau+\Omega\int_{0}^{t}H(t-\tau)a_{y}(\tau)d\tau\right)\nonumber \\
 & = & x(t)-\left[c_{1}+q_{x}\right]\nonumber \\
 & = & x(t)-{c_{1}}^{\prime}\nonumber \\
 & = & \int_{0}^{t}H_{0}(t-\tau)f_{x}(\tau)d\tau-\Omega^{2}\int_{0}^{t}H_{0}^{\prime}(t-\tau)f_{x}(\tau)d\tau+\Omega\int_{0}^{t}H(t-\tau)f_{y}(\tau)d\tau\;\;\;,\label{leqs}
\end{eqnarray}

\begin{eqnarray}
g_{2}(t) & = & y(t)-\left(A(t)y(0)+B(t)x(0)+C(t)u_{y}(0)-D(t)u_{x}(0)\right)\nonumber \\
 & - & \left(\int_{0}^{t}H_{0}(t-\tau)a_{y}(\tau)d\tau-\Omega^{2}\int_{0}^{t}H_{0}^{\prime}(t-\tau)a_{y}(\tau)d\tau-\Omega\int_{0}^{t}H(t-\tau)a_{x}(\tau)d\tau\right)\nonumber \\
 & = & y(t)-\left[c_{2}+q_{y}\right]\nonumber \\
 & = & y(t)-{c_{2}}^{\prime}\nonumber \\
 & = & \int_{0}^{t}H_{0}(t-\tau)f_{y}(\tau)d\tau-\Omega^{2}\int_{0}^{t}H_{0}^{\prime}(t-\tau)f_{y}(\tau)d\tau-\Omega\int_{0}^{t}H(t-\tau)f_{x}(\tau)d\tau\;\;\;,\label{leqs1}
\end{eqnarray}

\begin{eqnarray}
g_{3}(t) & = & u_{x}(t)-\left(\dot{A}(t)x(0)-\dot{B}(t)y(0)+\dot{C}(t)u_{x}(0)+\dot{D}(t)u_{y}(0)\right)\nonumber \\
 & - & \left(\int_{0}^{t}\dot{H_{0}}(t-\tau)a_{x}(\tau)d\tau-\Omega^{2}\int_{0}^{t}\dot{H_{0}^{\prime}}(t-\tau)a_{x}(\tau)d\tau+\Omega\int_{0}^{t}\dot{H}(t-\tau)a_{y}(\tau)d\tau\right)\nonumber \\
 & = & u_{x}(t)-\left[c_{3}+p_{x}\right]\nonumber \\
 & = & u_{x}(t)-{c_{3}}^{\prime}\nonumber \\
 & = & \int_{0}^{t}\dot{H_{0}}(t-\tau)f_{x}(\tau)d\tau-\Omega^{2}\int_{0}^{t}\dot{H_{0}^{\prime}}(t-\tau)f_{x}(\tau)d\tau+\Omega\int_{0}^{t}\dot{H}(t-\tau)f_{y}(\tau)d\tau\;\;\;,\label{leqs2}
\end{eqnarray}

\begin{eqnarray}
g_{4}(t) & = & u_{y}(t)-\left(\dot{A}(t)y(0)+\dot{B}(t)x(0)+\dot{C}(t)u_{y}(0)-\dot{D}(t)u_{x}(0)\right)\nonumber \\
 & - & \left(\int_{0}^{t}\dot{H_{0}}(t-\tau)a_{y}(\tau)d\tau-\Omega^{2}\int_{0}^{t}\dot{H_{0}^{\prime}}(t-\tau)a_{y}(\tau)d\tau-\Omega\int_{0}^{t}\dot{H}(t-\tau)a_{x}(\tau)d\tau\right)\nonumber \\
 & = & u_{y}(t)-\left[c_{4}+p_{y}\right]\nonumber \\
 & = & u_{y}(t)-{c_{4}}^{\prime}\nonumber \\
 & = & \int_{0}^{t}\dot{H_{0}}(t-\tau)f_{y}(\tau)d\tau-\Omega^{2}\int_{0}^{t}\dot{H_{0}^{\prime}}(t-\tau)f_{y}(\tau)d\tau-\Omega\int_{0}^{t}\dot{H}(t-\tau)f_{x}(\tau)d\tau\;\;\;,\label{leqs3}
\end{eqnarray}
where

\noindent 
\begin{equation}
A\equiv A(t)=\chi_{0}(t)+\Omega^{2}\omega^{2}\chi(t)\;\;,\label{leqs8}
\end{equation}

\begin{equation}
B\equiv B(t)=\Omega\omega^{2}H^{\prime}(t)\;\;,\label{leqs9}
\end{equation}
\begin{equation}
C\equiv C(t)=H_{0}(t)-\Omega^{2}H_{0}^{\prime}(t)\;\;,\label{leqs10}
\end{equation}

\begin{equation}
D\equiv D(t)=\Omega H(t)\;\;,\label{leqs11}
\end{equation}

\noindent with 
\begin{equation}
\chi_{0}(t)=1-\omega^{2}\int_{0}^{t}H_{0}(\tau)d\tau\;\;,\label{leqs12}
\end{equation}

\begin{equation}
\chi(t)=\int_{0}^{t}H_{0}^{\prime}(\tau)d\tau\;\;.\label{leqs13}
\end{equation}

\noindent The above Eqs.(\ref{leqs12}-\ref{leqs13}) implies that
$\chi_{0}(0)=1.0$ and $\chi(0)=0$. 

Now we have to define the functions $H_{0}(t)$, $H_{0}^{\prime}(t)$,
$H(t)$ and $H^{\prime}(t)$ which appear in the above equations.
These are the inverse Laplace transformation of $\tilde{H_{0}}(s)$,
$\tilde{H_{0}^{\prime}}(s)$, $\tilde{H}(s)$ and $\tilde{H^{\prime}}(s)$,
respectively. Here we have used $\tilde{H}(s)=s\tilde{H^{\prime}}(s)$
and $\tilde{H_{0}^{\prime}}(s)=s^{2}\tilde{H_{00}}(s)$. $\tilde{H_{0}}(s)$,
$\tilde{H^{\prime}}(s)$ and $\tilde{H_{00}}(s)$ are defined as

\begin{equation}
\tilde{H}_{0}(s)=\frac{1}{s^{2}+s\tilde{\gamma}(s)+\omega^{2}}\;\;,\label{leqs14}
\end{equation}

\begin{equation}
\tilde{H^{\prime}}(s)=\frac{1}{(s^{2}+s\tilde{\gamma}(s)+\omega^{2})^{2}+(\Omega s)^{2}}\;\;,\label{leqs15}
\end{equation}

\noindent and

\begin{equation}
\tilde{H_{00}}(s)=\frac{1}{(s^{2}+s\tilde{\gamma}(s)+\omega^{2})[(s^{2}+s\tilde{\gamma}(s)+\omega^{2})^{2}+(\Omega s)^{2}]}\;\;\;.\label{leqs16}
\end{equation}

\noindent Here $\tilde{\gamma}(s)$ is the Laplace transform of $\gamma(t)$. 

We now consider the fluctuations in position and velocity, respectively.
All the second moments corresponding to the fluctuations (as given
by Eqs.(\ref{leqs}-\ref{leqs3})) can be represented by the matrix,
$\boldsymbol{\mathcal{A}}^{\prime}(t)$ with $A_{ij}^{\prime}=\langle g_{i}(t)g_{j}(t)\rangle$.
Using Eqs.(\ref{leqs}-\ref{leqs3}), one can read the matrix $\boldsymbol{\mathcal{A}}^{\prime}(t)$
as 
\begin{equation}
\boldsymbol{\mathcal{A}}^{\prime}=\begin{bmatrix}A_{1} & 0 & A_{3} & A_{4}\\
0 & A_{1} & -A_{4} & A_{3}\\
A_{3} & -A_{4} & A_{2} & 0\\
A_{4} & A_{3} & 0 & A_{2}
\end{bmatrix}\label{eq78a}
\end{equation}
where we have used $A_{11}^{\prime}=A_{22}^{\prime}=A_{1}$, $A_{33}^{\prime}=A_{44}^{\prime}=A_{2}$,
$A_{13}^{\prime}=A_{31}^{\prime}=A_{24}^{\prime}=A_{42}^{\prime}=A_{3}$,
$A_{14}^{\prime}=A_{41}^{\prime}=A_{4}$, and $A_{23}^{\prime}=A_{32}^{\prime}=-A_{4}$.
Then the inverse of the above matrix can be written as 
\begin{equation}
{\boldsymbol{\mathcal{A}}^{\prime}}^{-1}(t)=\frac{adj(\boldsymbol{\mathcal{A}}^{\prime}(t))}{|\boldsymbol{\mathcal{A}}^{\prime}(t)|}=\frac{1}{(A_{1}A_{2}-A_{3}^{2}-A_{4}^{2})}\begin{bmatrix}A_{2} & 0 & -A_{3} & -A_{4}\\
0 & A_{2} & A_{4} & -A_{3}\\
-A_{3} & A_{4} & A_{1} & 0\\
-A_{4} & -A_{3} & 0 & A_{1}
\end{bmatrix}
\end{equation}

\section{Time-dependent coefficients of the Fokker-Planck equation (\ref{eq84})}

\noindent Following Ref.\cite{arxiv} we use the distribution function (\ref{eq76}) in Eq. (\ref{eq84}) and then get the time dependent  coefficients as

\noindent 
\begin{equation}
H_{1}=\frac{\left[-\ddot{A}\tilde{a}_{x}(t)-\ddot{B}\tilde{a}_{y}(t)-\ddot{C}\tilde{a}_{v_{x}}(t)-\ddot{D}\tilde{a}_{v_{y}}(t))\right]}{\Delta_{m}},\label{h1}
\end{equation}
\begin{equation}
H_{2}=\frac{\left[\ddot{A}\tilde{b}_{x}(t)-\ddot{B}\tilde{b}_{y}(t)-\ddot{C}\tilde{b}_{v_{x}}(t)-\ddot{D}\tilde{b}_{v_{y}}(t))\right]}{\Delta_{m}},\label{h2}
\end{equation}
\begin{equation}
H_{3}=\frac{\left[-\ddot{A}\tilde{d}_{x}(t)+\ddot{B}\tilde{d}_{y}(t)-\ddot{C}\tilde{d}_{v_{x}}(t)+\ddot{D}\tilde{d}_{v_{y}}(t))\right]}{\Delta_{m}},\label{h3}
\end{equation}

\begin{equation}
H_{4}=\frac{\left[-\ddot{A}\tilde{c}_{x}(t)+\ddot{B}\tilde{c}_{y}(t)+\ddot{C}\tilde{c}_{v_{x}}(t)-\ddot{D}\tilde{c}_{v_{y}}(t))\right]}{\Delta_{m}},\label{h6}
\end{equation}
\begin{equation}
H_{5}=\left[\dot{A_{4}}+H_{2}A_{4}+H_{3}A_{3}-H_{4}A_{1}\right],\label{h7}
\end{equation}
\begin{equation}
H_{6}=\left[\dot{A_{3}}-A_{2}-H_{3}A_{4}+H_{1}A_{1}+H_{2}A_{3}\right],\label{h8}
\end{equation}

\begin{equation}
H_{7}=\frac{1}{2}\left[\dot{A_{2}}+2H_{1}A_{3}+2H_{2}A_{2}-2H_{4}A_{4}\right],\label{h10}
\end{equation}
Here we have used

\noindent 
\begin{equation}
\Delta_{m}=(A^{2}+B^{2})(\dot{C}^{2}+\dot{D}^{2})+(C^{2}+D^{2})(\dot{A}^{2}+\dot{B}^{2})-2(AC-BD)(\dot{A}\dot{C}-\dot{B}\dot{D})-2(AD+BC)(\dot{A}\dot{D}+\dot{B}\dot{C})\;\;\;,\label{del}
\end{equation}

\[
\tilde{a}_{x}(t)=A(\dot{C}^{2}+\dot{D}^{2})-C(\dot{A}\dot{C}-\dot{B}\dot{D})-D(\dot{A}\dot{D}+\dot{B}\dot{C}),
\]

\[
\tilde{c}_{x}(t)=B(\dot{C}^{2}+\dot{D}^{2})+D(\dot{A}\dot{C}-\dot{B}\dot{D})-C(\dot{A}\dot{D}+\dot{B}\dot{C}),
\]

\[
\tilde{b}_{x}(t)=B(C\dot{D}-\dot{C}D)+C(A\dot{C}-\dot{A}C)+D(A\dot{D}-D\dot{A}),
\]

\begin{equation}
\tilde{d}_{x}(t)=B(C\dot{C}+D\dot{D})-C(A\dot{D}+C\dot{B})+D(A\dot{C}-D\dot{B}).\label{abcdx}
\end{equation}

\[
\tilde{a}_{y}(t)=B(\dot{C}^{2}+\dot{D}^{2})+D(\dot{A}\dot{C}-\dot{B}\dot{D})-C(\dot{A}\dot{D}+\dot{B}\dot{C}),
\]

\[
\tilde{c}_{y}(t)=A(\dot{C}^{2}+\dot{D}^{2})-C(\dot{A}\dot{C}-\dot{B}\dot{D})-D(\dot{A}\dot{D}+\dot{B}\dot{C}),
\]

\[
\tilde{b}_{y}(t)=A(C\dot{D}-\dot{C}D)-C(B\dot{C}-\dot{B}C)-D(B\dot{D}-\dot{B}D),
\]

\begin{equation}
\tilde{d}_{y}(t)=A(C\dot{C}+\dot{D}D)+C(B\dot{D}-\dot{A}C)-D(B\dot{C}+\dot{A}D).\label{abcdy}
\end{equation}

\[
\tilde{a}_{v_{x}}(t)=C(\dot{A}^{2}+\dot{B}^{2})-A(\dot{A}\dot{C}-\dot{B}\dot{D})-B(\dot{A}\dot{D}+\dot{B}\dot{C}),
\]

\[
\tilde{c}_{v_{x}}(t)=D(\dot{A}^{2}+\dot{B}^{2})+B(\dot{A}\dot{C}-\dot{B}\dot{D})-A(\dot{A}\dot{D}+\dot{B}\dot{C}),
\]

\[
\tilde{b}_{v_{x}}(t)=A(A\dot{C}-\dot{A}C)+B(B\dot{C}-\dot{B}C)-D(A\dot{B}-\dot{A}B),
\]

\begin{equation}
\tilde{d}_{v_{x}}(t)=A(A\dot{D}+\dot{B}C)+B(B\dot{D}-\dot{A}C)-D(A\dot{A}+\dot{B}B).\label{abcdvx}
\end{equation}

\[
\tilde{a}_{v_{y}}(t)=D(\dot{A}^{2}+\dot{B}^{2})+B(\dot{A}\dot{C}-\dot{B}\dot{D})-A(\dot{A}\dot{D}+\dot{B}\dot{C}),
\]

\[
\tilde{c}_{v_{y}}(t)=C(\dot{A}^{2}+\dot{B}^{2})-A(\dot{A}\dot{C}-\dot{B}\dot{D})-B(\dot{A}\dot{D}+\dot{B}\dot{C}),
\]

\[
\tilde{b}_{v_{y}}(t)=A(A\dot{D}-\dot{A}D)+B(B\dot{D}-\dot{B}D)+C(A\dot{B}-\dot{A}B),
\]

\noindent 
\begin{equation}
\tilde{d}_{v_{y}}(t)=A(A\dot{C}-\dot{B}D)+B(B\dot{C}+\dot{A}D)-C(A\dot{A}+\dot{B}B).\label{abcdvy}
\end{equation}

\noindent It is to be noted here that all the time-dependent coefficients
(as given by Eqs.(\ref{h1}-\ref{h10})) do not depend on $c_{1}$, $c_{2}$, $c_{3}$and $c_{4}$ as like
as the previous cases with $\boldsymbol{a}(t)=0$. Then following
the earlier cases we have

\begin{equation}
G_{1}=\dot{p_{x}}+H_{1}q_{x}+H_{2}p_{x}-H_{3}p_{y}+H_{4}q_{y},\label{g1}
\end{equation}

\begin{equation}
G_{2}=\dot{p_{y}}+H_{1}q_{y}+H_{2}p_{y}+H_{3}p_{x}-H_{4}q_{x},\label{g2}
\end{equation}

\end{document}